\documentclass[aps,prb,twocolumn,groupedaddress,showpacs]{revtex4}
\usepackage{graphicx}

\begin{document}


\title{Proximity-Induced Superconductivity in Platinum Metals}


\author{D. Katayama}
\author{A. Sumiyama}
\email[Corresponding Author: ]{sumiyama@sci.himeji-tech.ac.jp}
\author{Y. Oda}
\affiliation{Faculty of Science, Himeji Institute of Technology, Kamigori-cho, Ako-gun, Hyogo 678-1297, Japan}


\date{\today}

\begin{abstract}
The diamagnetism of platinum metals (N: Rh, Pt, Pd), which is induced by the proximity effect of a superconductor (S: Nb), has been investigated for N-S double layers. Notwithstanding the strong spin fluctuation in platinum metals, the screening distance $\rho$ in N increases with a decrease in temperature and reaches a value which is expected in comparison with $\rho$ in Cu. When magnetic impurities are included in N, the proximity effect is drastically suppressed and the paramagnetism due to a giant moment is observed. 
\end{abstract}


\maketitle

The investigation of superconductivity in platinum metals (PM) has  been of considerable interest over the last few decades. Despite their high electronic specific-heat coefficient which favors superconductivity, Pt and Pd have not become superconductive, while Rh is the element which has the lowest transition temperature $T_{{\rm c}}$=325 $\mu$K.\cite{1} It is generally agreed that spin fluctuation (paramagnon) effect reduces $T_{{\rm c}}$ in these three elements; although both electron-phonon interaction and spin fluctuation contribute to the mass enhancement, they play an opposite role in the occurrence of superconductivity.\cite{2,3} Spin fluctuations are reflected also in the strongly exchange-enhanced paramagnetism of PM. The recent observation of the superconductivity in Pt powder is ascribed to the reduction of the spin fluctuation by the spin-orbit scattering at rough surfaces.\cite{4,5} Tunneling study of Pd, however, indicates that the paramagnon effect is less important for the absence of superconductivity.\cite{6}

In order to clarify how the spin fluctuation affect superconductivity, it will be useful to introduce the Cooper pairs to PM by the proximity effect of an adjacent superconductor, and observe the destruction of them. In addition, the proximity effect in PM may be useful to reveal the difference between the BCS (singlet) and the triplet superconductivity; if the Cooper pairs are introduced from the triplet superconductors, such as UPt$_{3}$ or Sr$_{2}$RuO$_{4}$, they are thought to be less sensitive to the spin fluctuation effect.
  
Proximity-induced superconductivity of a normal metal (N) has been investigated through the measurement of the diamagnetic response of N-clad S wires, where S is a superconductor and noble metals (Cu, Au, Ag) are used as N.\cite{7,8,9} In contrast to PM, the absence of superconductivity in these noble metals are attributed to the weak electron-phonon interaction and the low electronic specific-heat coefficients. As for PM, there exists one report that the proximity effect in Pd-clad Nb wire is not observed.\cite{9} The leakage of the Cooper pairs, however, is so sensitive to the quality of the N-S interface that further work on different type of samples is needed.

Recently, we have reported the proximity effect of N-S double layers.\cite{10} In N-clad S wires, an N-S interface is obtained during a wire-drawing process, so that post-annealing, which may degrade the N-S interface, is needed to reduce mechanical imperfections in the lattice of N. In our N-S double layers, on the other hand, S is deposited on N which has been already annealed. This process enables us to anneal N at high temperatures to improve the electronic mean free path $\ell_{{\rm N}}$ without causing damage to the N-S interface. In this paper, the diamagnetic response of PM (Rh, Pt, Pd)-Nb double layers is described and is discussed from the viewpoint of the spin fluctuation and the electron-phonon interaction in PM. 

%
\begin{table}
 \caption{\label{table1}Properties of commercial platinum metal sheets}
 \begin{ruledtabular}
 \begin{tabular}{lrl}
 Sheet&$d$ ($\mu$m)&Major impurities (wt. ppm)\\
 \hline
 Rh(3N)\footnotemark[1]&100&Pt 154; Si 114; Fe 82; Ir 77; Cr 39; Cd 38\\
 Pt(4N)\footnotemark[1]&50&Si 20; Fe 16; Mg 5; Rh 5; Pd 1; Ag 1\\
 Pt(5N)\footnotemark[2]&100&Rh 5; Ir 4; Pd 2; Al 2\\
 Pd(4N)\footnotemark[1]&100&Pt 21; Si 20; Fe 17; Au 1; Cu 1; Ag 1\\
  
 \end{tabular}
 \end{ruledtabular}
 \footnotetext[1]{Furuuchi Chemical}
 \footnotetext[2]{Johnson Matthey}
 \end{table}
%
The N-S double layers were fabricated by use of commercial platinum metal sheets (Pd, Pt, Rh), of which source, purity, thickness $d$, and major impurities are listed in Table\ \ref{table1}. The sheets whose thickness is 100 $\mu$m were rolled out to be $d_{{\rm N}}$=50 $\mu$m. All the sheets were cut up into strips 1 mm wide and 10mm long. The strips were annealed for one hour to remove the effect of cold work. The details in annealing conditions are described in Table\ \ref{table2}. The residual resistance ratio $RRR$ between room temperature and 4.2 K, which is determined by resistance measurements along strips, is tabulated also. The Cu(4N) sample is the one used in our previous investigation.\cite{10} 

\begin{figure}[b]
\begin{center}
\includegraphics*[width=1.0\linewidth, trim=2cm 9cm 2cm 9cm, clip]{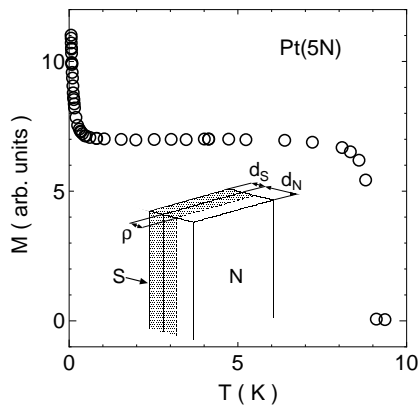}
\caption{\label{fig1}Typical temperature dependence of the mutual inductance $M$ measured in arbitrary units. The inset shows schematic of N-S double layers. The shadowed area displays the Meissner effect.}
\end{center}
\end{figure}

%
\begin{table}
 \caption{\label{table2}Properties of N in N-S double layers}
 \begin{ruledtabular}
 \begin{tabular}{lrrc}
 Sample&Annealing&$RRR$&$\xi_{{\rm N0}} \cdot \sqrt{T}$ ($\mu$m$\cdot \sqrt{{\rm K}}$)\\
 \hline
 Rh(3Na)&1200$^{\circ }$C (in air)&260&0.52\\
 Rh(3Nb)&800$^{\circ }$C (in Ar)&60&0.25\\
 Pt(4N)&600$^{\circ }$C (in Ar)&110&0.20\\
 Pt(5N)&600$^{\circ }$C (in Ar)&940&0.60\\
 Pd(4N)&500$^{\circ }$C (in Ar)&99&0.16\\
 Cu(4N)&600$^{\circ }$C (in Ar)&120&1.5\\
  
 \end{tabular}
 \end{ruledtabular}
 \end{table}
The surface of the PM strips was rf sputter cleaned by Ar ion, and then a Nb layer, of which thickness $d_{{\rm S}}$=12 $\mu$m, was deposited by rf sputtering technique, as shown in the inset of Fig. \ref{fig1}. The strips were held at room temperature during the deposition process. Hereafter, the N-S double layers are called, for example, "Rh(3Na)", where "3N" denotes the purity and "a" denotes the different annealing condition.
 
The N-S double layers were electrically insulated by varnish, and a bundle of about 30 strips were mounted in a mutual inductance coil of a Hartshorn bridge in parallel to the magnetic field. It was linked to the mixing chamber of a dilution refrigerator and cooled down to 50 mK. All measurements were performed at a frequency of 280 Hz in an ac field as low as 6 mOe. No frequency dependence was observed between 40 and 280 Hz. The earth magnetic field was reduced to a few mOe by a $\mu$-metal shield. The temperature was determined using the carbon thermometers which were calibrated by a CMN thermometer.

%
 Figure \ref{fig1} shows typical results for the temperature dependence of the mutual inductance $M$ of the coil when the Pt(5N) sample was mounted. As the temperature is decreased, the change in $M$ due to the superconducting transition of the Nb layer, followed by a nearly constant $M$, and the further change due to the proximity-induced diamagnetism of Pt is observed.
 
%
\begin{figure}[t]
\begin{center}
\includegraphics*[width=1.0\linewidth, trim=2cm 10cm 2cm 4cm, clip]{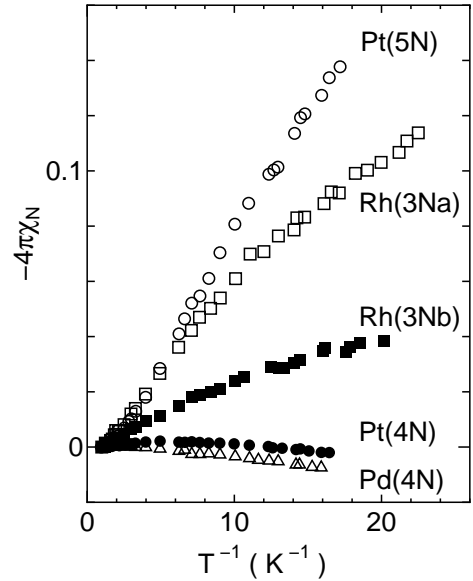}
\caption{\label{fig2}Temperature dependence of the ac magnetic susceptibility $\chi_{{\rm N}}$ of platinum metals (N) for five N-S double layers.}
\end{center}
\end{figure}
On the assumption that the Nb layer shows full diamagnetism ($\chi_{{\rm S}}=-1/4\pi$), the susceptibility $\chi_{{\rm N}}$ of N is given by
\begin{equation}
\chi_{{\rm N}}=-\frac{1}{4\pi }\frac{d_{{\rm S}}\Delta M_{{\rm N}}}{d_{{\rm N}}\Delta M_{{\rm S}}},
\label{eq1}
\end{equation}
where $\Delta M_{{\rm S}}$ and $\Delta M_{{\rm N}}$ are the mutual inductance change due to the superconducting transition of the Nb layer and the magnetism in N, respectively. Although the transition of the Nb layer and the proximity effect in N successively occur, the change in $M$ is found to be small between 1 K and 5 K for the whole samples, so that we assume that the change in $\chi_{{\rm N}}$ appears below 1 K, and take $\Delta M_{{\rm S}}$=$M$(9.5 K)-$M$(1 K) and $\Delta M_{{\rm N}}$=$M$(1 K)-$M$($T$). 

We show in Fig. \ref{fig2} the temperature dependence of  $\chi_{{\rm N}}$ of the whole samples, which is plotted as $-4\pi \chi_{{\rm N}}$ vs. $T^{-1}$. The proximity-induced diamagnetism is observed for Pt(5N) and Rh(3N), while Pd(4N) and Pt(4N) show a small paramagnetic signal at low temperatures. The difference between Pt(5N) and Pt(4N) suggests that the magnetic impurities, which consist mainly of Fe, play an important role in the absence of the proximity effect. Although Rh contains the largest amount of Fe impurities, it is reported that Fe in Rh does not display a Kondo effect.\cite{11}

%
\begin{figure}[b]
\begin{center}
\includegraphics*[width=1.0\linewidth, trim=0cm 8.5cm 0cm 11cm, clip]{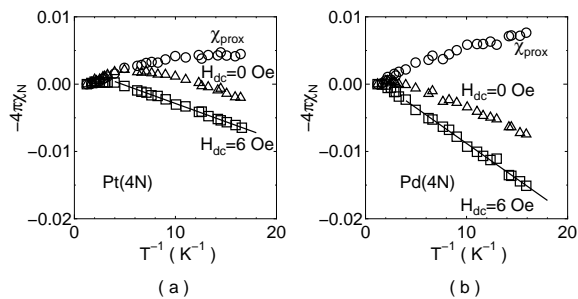}
\caption{\label{fig3}Temperature dependence of the ac magnetic susceptibility $\chi_{{\rm N}}$ for Pt(4N) and Pd(4N) when a dc magnetic field of 6 Oe is applied or not. The difference $\chi_{{\rm prox}}=\chi_{{\rm N}}(0\ {\rm Oe})-\chi_{{\rm N}}(6\ {\rm Oe})$ is ascribed to the proximity-induced diamagnetism. The solid lines indicate a least square fit to the lower-temperature data using the Curie law.}
\end{center}
\end{figure}
In Pt(4N) and Pd(4N), the diamagnetic susceptibility $\chi_{{\rm prox}}$ due to the proximity effect is expected to be small, so that the paramagnetic susceptibility $\chi_{{\rm imp}}$ due to the magnetic impurities should be taken into account; $\chi_{{\rm N}}$ is expressed as $\chi_{{\rm N}}=\chi_{{\rm prox}}+\chi_{{\rm imp}}$. Since the proximity effect is suppressed by applying a small field as low as a few Oe,\cite{7} $\chi_{{\rm imp}}$ can be evaluated by measuring $\chi_{{\rm N}}$ in dc magnetic field $H_{{\rm dc}}$=6 Oe, as shown in Fig. \ref{fig3}.  The difference $\chi_{{\rm prox}}=\chi_{{\rm N}}(0\ {\rm Oe})-\chi_{{\rm N}}(6\ {\rm Oe})$ is plotted also. The fact that the change in $\chi_{{\rm N}}$ by applying $H_{{\rm dc}}$=6 Oe is ascribed to the proximity effect is confirmed also by the absence of  $H_{{\rm dc}}$ dependence of $\chi_{{\rm N}}$ in another Pd(4N) sample, in which an insulating SiO$_{2}$ layer 1 $\mu$m in thickness is inserted between Nb and Pd.

Since $\chi_{{\rm N}}$(6 Oe) increases approximately in proportion to $T^{-1}$ at low temperatures, we fit the data to the Curie law, as indicated by the solid lines in Fig. \ref{fig3}. We obtain the effective Bohr magneton of 7.5 $\mu_{{\rm B}}$ and 14 $\mu_{{\rm B}}$ for Pt and Pd, respectively, if we use the impurity levels of Fe in Table\ \ref{table1}, which are given by the suppliers. These values agree well with the reported ones of "giant magnetic moments": 8 $\mu_{{\rm B}}$ in Pt and 13-16 $\mu_{{\rm B}}$ in Pd.\cite{12}

%
\begin{figure}[b]
\begin{center}
\includegraphics*[width=1.0\linewidth, trim=0cm 9.5cm 0cm 7.5cm, clip]{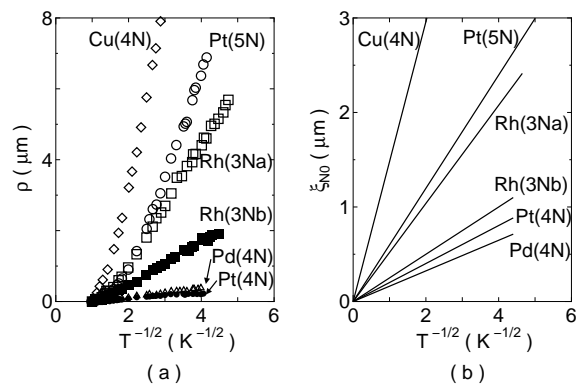}
\caption{\label{fig4}Temperature dependence of the screening distance $\rho$ and the  coherence length $\xi_{{\rm N0}}$. The $\rho$ values of  Pt(4N) and Pd(4N) are calculated from $\chi_{{\rm prox}}$ in Fig. \ref{fig3}, and $\xi_{{\rm N0}}$ is calculated using the relation in Table \ref{table2}.}
\end{center}
\end{figure}
The proximity effect in Pt(4N) and Pd(4N), which is derived from $H_{{\rm dc}}$ dependence of $\chi_{{\rm N}}$, is extremely small as compared with the other samples. In Fig. \ref{fig4}(a), the screening distance $\rho$ of the magnetic field in normal metals, which is expressed as $\rho=-4\pi \chi_{{\rm prox}}d_{{\rm N}}$ for Pd(4N) and Pt(4N), or $\rho=-4\pi \chi_{{\rm N}}d_{{\rm N}}$ for the other samples, is plotted as a function of $T^{-1/2}$.

The theoretical derivation of $\rho$ was first made by de Gennes and his co-workers, as given by
\begin{equation}
\rho=\xi_{{\rm N}}(T)\{\ln(\xi_{{\rm N}}(T)/\lambda_{{\rm N}}(T))-0.116\},
\label{eq3}
\end{equation}
where $\lambda_{{\rm N}}(T)$ is the penetration depth in the normal metal N at the interface and $\xi_{{\rm N}}(T)$ is the coherence length in N.\cite{13} In the dirty case where the electronic mean-free path $\ell_{{\rm N}}$ in N is shorter than $\xi_{{\rm N}}(T)$, $\xi_{{\rm N}}(T)$ is expressed as 
\begin{equation}
\xi_{{\rm N}}(T)=\sqrt{\frac{\hbar v_{{\rm N}}\ell_{{\rm N}}}{6\pi k_{{\rm B}}T}}\left(1-\frac{2N_{{\rm N}}V_{{\rm N}}}{1-CN_{{\rm N}}V_{{\rm N}}}\right)^{-1/2},
\label{eq4}
\end{equation}
where $v_{{\rm N}}$ is the Fermi velocity, $N_{{\rm N}}$ is the electron density at the Fermi level, and $V_{{\rm N}}$ is the electron-electron interaction in N; $V_{{\rm N}}$ is either positive (attractive) or negative (repulsive). The variable $C$ is given by $C=\ln (1.14\theta_{{\rm D}}/T)-2$, where $\theta_{{\rm D}}$ is the Debye temperature of N.\cite{13'} Since $\lambda_{{\rm N}}(T)$ is proportional to $\sqrt{T}/F_{{\rm N}}(T)$, where $F_{{\rm N}}(T)$ is the condensation amplitude in N at the interface,\cite{13} $\rho$ becomes measurable when $F_{{\rm N}}(T)$ grows to be sufficiently large with decreasing temperatures below $T_{{\rm c}}$. At low temperatures, $\ln(\xi_{{\rm N}}(T)/\lambda_{{\rm N}}(T))$ increases only slowly, so that $\rho$ shows the same temperature dependence as $\xi_{{\rm N}}(T)$.

In the case of $N_{{\rm N}}V_{{\rm N}}=0$, $\xi_{{\rm N}}(T)$  is calculated by the equation
\begin{equation}
\xi_{{\rm N0}}(T)=\sqrt{\frac{\hbar \pi k_{{\rm B}}}{6e^{2}\gamma \rho_{0}}}\times T^{-1/2},
\label{eq5}
\end{equation}
where $\gamma$ is the linear specific heat coefficient and $\rho_{0}$ is the electrical resistivity, which is calculated from the residual resistance ratio $RRR$ in Table\ \ref{table2}.\cite{14} The results are listed in Table\ \ref{table2} and indicated by the solid lines in Fig. \ref{fig4}(b).

Except for Pt(4N) and Pd(4N), it is obvious that $\rho$ is proportional to $T^{-1/2}$, and the magnitude of $\rho$ reflects $\xi_{{\rm N0}}(T)$ directly; the difference in $\ln(\xi_{{\rm N}}(T)/\lambda_{{\rm N}}(T))$ in eq. (\ref{eq3}) is thought to be small among these samples. Since $\lambda_{{\rm N}}(T)$ is a decreasing function of the condensation amplitude in N at the interface, the present result indicates that the leakage of the Cooper pairs through the Pt-Nb and Rh-Nb interfaces is comparable to the Cu-Nb interface, and gives some evidence that our method is useful to obtain a clean N-S interface.

Since we have not investigated $\rho$ in Pd without magnetic impurities yet, we can not dismiss the possibility that the suppression of the proximity effect in Pd(4N) is ascribed to the degraded interface between Pd and Nb. Still, the order-of-magnitude agreement of $\rho$ between Pd(4N) and Pt(4N), which contain similar amount of Fe impurity, suggests that the $\rho$ value is decreased significantly by the magnetic impurities also in Pd(4N). When N contains magnetic impurities, $\xi_{{\rm N}}(T)$ is expressed as
\begin{equation}
\xi_{{\rm Nmag}}(T)=\sqrt{\frac{\hbar v_{{\rm N}}\ell_{{\rm N}}}{6\pi k_{{\rm B}}(T+\hbar /\pi k_{{\rm B}} \tau_{{\rm s}})}},
\label{eq6}
\end{equation}
where $1/\tau_{{\rm s}}$ is the exchange scattering rate of the electrons from the magnetic impurities.\cite{15} The reduction of the coherence length, which is given as $\xi_{{\rm Nmag}}/\xi_{{\rm N0}}=(1+\hbar/\pi k_{{\rm B}} T\tau_{{\rm s}})^{-1/2}$, becomes, for example, 0.16 at T=60 mK and with $\tau_{{\rm s}}=10^{-12}$ s, and explains the decrease in $\rho$ in Pd(4N) and Pt(4N), at least, qualitatively. Similar reduction of $\rho$ in Cu doped with magnetic impurities has been reported in our previous paper.\cite{16}

Although the spin fluctuation in PM increases the moment of the magnetic impurities and probably enhances the pair-breaking effect, the results for Pt(5N) and Rh(3N) suggest that the spin fluctuation has little effect on the proximity effect without magnetic impurities. This may be explained by the fact that the spin fluctuation leads to an enhanced singlet-state repulsion\cite{17} and the electron-phonon interaction probably balances it out; the total electron-electron coupling, whether positive or negative, is so small in eq. (\ref{eq4}), and the temperature range where the present measurements are performed is so high that neither decrease in $\rho$ nor deviation from the $T^{-1/2}$ dependence has been observed. Considering that the pair breaking effect by the spin fluctuation alone is not observed in PM in contact with a BCS (singlet) superconductor,  it seems that PM is not suitable for investigating the difference between a singlet and a triplet superconductor.

In summary, the suppression of the proximity-induced diamagnetism in platinum metals, which is ascribed to the spin-fluctuation effect, has not been observed. This result may reflect the small electron-electron coupling in them, which consists of the elctron-phonon (attractive) interaction and the repulsive interaction due to the spin-fluctuation. In Pt and Pd which include magnetic impurities, the spin fluctuation enhances their moment, and the pair-breaking effect reduces the proximity effect significantly. 
\begin{acknowledgments}
We would like to thank Dr. Y. Hasegawa for helpful discussions. This work was supported partly by a grant-in-aid from the MEXT, Japan.
\end{acknowledgments}
%
%

\end{document}